\documentstyle[dina4p,12pt]{article}

\title{ Numerical simulation of heavy fermions in
        an $\rm SU(2)_L \otimes SU(2)_R$ symmetric Yukawa model}

\author{
 {C. Frick}\thanks{Institut f\"ur Theoretische Physik E,
  RWTH Aachen, D-5100 Aachen, FRG,
  and HLRZ at KFA J\"ulich, D-5170 J\"ulich, FRG}
\and
 {L. Lin}\thanks{Institut f\"ur Theoretische Physik I,
  Universit\"at M\"unster, Wilhelm-Klemm-Str.~9, D-4400 M\"unster, FRG}
 \and
 {I. Montvay}\thanks{Deutsches Elektronen-Synchrotron DESY,
                 Notkestr.\,85, D-2000 Hamburg 52, FRG}
\and
 \addtocounter{footnote}{-1}%
 {$\mbox{G. M\"unster}^{\scriptstyle
 \fnsymbol{footnote} }$}
\and
 {$\mbox{M. Plagge}^{\scriptstyle
 \fnsymbol{footnote} }$}
\and
 \addtocounter{footnote}{-1}%
 {$\mbox{T. Trappenberg}^{\scriptstyle
 \fnsymbol{footnote} }$}
\and
 \addtocounter{footnote}{+2}%
 {$\mbox{H. Wittig}^{\scriptstyle
 \fnsymbol{footnote} }$}
 }

\newcommand{\be}{\begin{equation}}
\newcommand{\ee}{\end{equation}}
\newcommand{\half}{\frac{1}{2}}

\begin{document}
\maketitle

\begin{abstract} \normalsize
 An exploratory numerical study of the influence of heavy fermion
 doublets on the mass of the Higgs boson is performed in the decoupling
 limit of a chiral $\rm SU(2)_L \otimes SU(2)_R$ symmetric Yukawa model
 with mirror fermions.
 The behaviour of fermion and boson masses is investigated at infinite
 bare quartic coupling on $4^3 \cdot 8$, $6^3 \cdot 12$ and
 $8^3 \cdot 16$ lattices.
 A first estimate of the upper bound on the renormalized quartic
 coupling as a function of the renormalized Yukawa-coupling is given.
\end{abstract}
\vspace{1cm}


\section{ Introduction }                                    \label{s1}

 Recent LEP measurements \cite{LEP} fix the number of light neutrinos
 to three, therefore a simple further repetition of fermion families
 is excluded.
 Extensions of the minimal Standard Model by heavy fermions are,
 however, possible.
 Examples are: a fourth fermion family with heavy neutrino (for a
 recent reference see \cite{MARCI}), or a duplication of the three
 light families by heavy mirror families \cite{MIRFAM,ROMAPR}.
 Some limitations on the number of heavy fermions follow from
 studies of 1-loop radiative corrections \cite{PESTAK,ALTBAR}
 because of the non-decoupling of heavy fermions.
 The question of non-decoupling in higher loop orders is, however,
 open.
 In fact, one of the goals of lattice studies is to investigate this
 in the nonperturbative regime of couplings.

 The lattice formulation of the electroweak Standard Model is
 difficult because of the fermion doublers \cite{NIENIN}.
 In fact, at present no completely satisfactory formulation is known
 \cite{TSUKPR}: if one insists on explicit chiral gauge invariance,
 then mirror fermion fields have to be introduced \cite{CHFER},
 otherwise one has to fix the gauge as in the ``Rome-approach''
 \cite{ROMA2}.
 (In a recently proposed method \cite{KAPLAN} a fifth extra dimension
  has to be introduced.)

 There is, however, an interesting limit of the Standard Model
 which can be numerically simulated by present techniques.
 Namely, if the $\rm SU(3)_{colour} \otimes U(1)_{hypercharge}$
 gauge couplings are neglected, then, as a consequence of the
 pseudo-reality of SU(2) representations, mirror fermions can be
 transformed to normal fermions by charge conjugation, and Yukawa models
 with an even number ($N_f$) of degenerate fermion doublets can be
 simulated \cite{TSUKPR}.
 (For instance, $N_f=4$ corresponds to a heavy degenerate fermion
 family.)
 This can be done at nonzero Yukawa-couplings for both fermion
 ($G_\psi$) and mirror fermion ($G_\chi$), or by keeping $G_\chi$ and
 the fermion-mirror-fermion mixing mass $\mu_{\psi\chi}$
 at zero, and thereby decoupling the mirror fermion exactly from
 the physical spectrum \cite{ROMA1}.

 In the present paper we choose this second way, where the exact
 decoupling in the continuum limit is assured by the Golterman-Petcher
 fermion shift symmetry \cite{GOLPET}.
 This symmetry is exact at $G_\chi=\mu_{\psi\chi}=0$, and implies
 a set of identities, which makes the parameter tuning easier.
 Another interpretation of the decoupling limit also deserves attention.
 Namely, interchanging the r\^oles of fermion and mirror fermion by
 considering $\chi$ to be the fermion and $\psi$ the mirror fermion, the
 decoupling scenario turns out to be a rather good approximation of the
 situation in phenomenological models with mirror fermions
 \cite{MIRFAM}.
 This is due to the fact that all known physical fermions have very
 small Yukawa-couplings.
 The only fermion states with strong Yukawa-couplings would be the
 members of the mirror families, if they would exist.
 In fact, the smallness of the known fermion masses on the electroweak
 scale could then be explained by the approximate validity of the
 Golterman-Petcher shift symmetry.

 An important set of questions for the numerical simulations is
 concentrated around the ``allowed range of renormalized couplings'',
 which is cut-off dependent and shrinks to zero for infinite cut-off
 if the continuum limit is trivial.
 It is expected on the basis of 1-loop perturbation theory that,
 as a function of the renormalized Yukawa-coupling, the allowed
 region for the renormalized quartic coupling is limited by an
 upper bound obtained at infinite bare quartic coupling
 ($\lambda=\infty$), and by a lower bound called ``vacuum stability
 bound'' reached at zero bare quartic coupling ($\lambda=0$)
 (for discussions see \cite{FKLMMM,LMMW,TSUKPR}).
 In the present work the numerical simulations are restricted to
 $\lambda=\infty$, and first results on the behaviour of the upper bound
 are obtained.
 The study of the $\lambda \to 0$ limit is postponed to future work.


\section{ Lattice action, decoupling, physical quantities } \label{s2}

 Our numerical simulations were performed in the chiral
 $\rm SU(2)_L \otimes SU(2)_R$ symmetric Yukawa model with
 $N_f = 2$ mirror pairs of fermion doublet fields in the decoupling
 limit.
 The conventions in the lattice action and the definition of different
 renormalized physical quantities closely follow our previous papers
 on $\rm U(1)_L \otimes U(1)_R$ \cite{FKLMMM,LMMW} and
 $\rm SU(2)_L \otimes SU(2)_R$ \cite{LINWIT} symmetric models.
 Therefore we only repeat here the most essential formulae, and
 include the definitions specific to the present investigation.

 The lattice action is a sum of the O(4)
 ($\cong \rm SU(2)_L \otimes SU(2)_R$) symmetric pure scalar part
 $S_\varphi$ and fermionic part $S_\Psi$:
\be \label{eq01}
S = S_\varphi + S_\Psi \ .
\ee
 $\varphi_x$ is the $2 \otimes 2$ matrix scalar field, and
 $\Psi_x \equiv (\psi_x, \chi_x)$ stands for the mirror pair of fermion
 doublet fields (usually $\psi$ is the fermion doublet and $\chi$ the
 mirror fermion doublet).
 In the usual normalization conventions for numerical simulations we
 have
$$
S_\varphi = \sum_x \left\{ \half {\rm Tr\,}(\varphi_x^+\varphi_x) +
\lambda \left[ \half{\rm Tr\,}(\varphi_x^+\varphi_x) - 1\right]^2
- \kappa\sum_{\mu=1}^4
{\rm Tr\,}(\varphi^+_{x+\hat{\mu}}\varphi_x)
\right\} \ ,
$$
$$
S_\Psi = \sum_x \left\{ \mu_{\psi\chi} \left[
(\overline{\chi}_x\psi_x) + (\overline{\psi}_x\chi_x) \right]
\right.
$$
$$
- K \sum_{\mu=\pm 1}^{\pm 4} \left[
(\overline{\psi}_{x+\hat{\mu}} \gamma_\mu \psi_x) +
(\overline{\chi}_{x+\hat{\mu}} \gamma_\mu \chi_x)
+ r \left( (\overline{\chi}_{x+\hat{\mu}}\psi_x)
- (\overline{\chi}_x\psi_x)
+ (\overline{\psi}_{x+\hat{\mu}}\chi_x)
- (\overline{\psi}_x\chi_x)  \right) \right]
$$
\be \label{eq02}
\left.
+ G_\psi \left[ (\overline{\psi}_{Rx}\varphi^+_x\psi_{Lx}) +
(\overline{\psi}_{Lx}\varphi_x\psi_{Rx}) \right]
+ G_\chi \left[ (\overline{\chi}_{Rx}\varphi_x\chi_{Lx}) +
(\overline{\chi}_{Lx}\varphi^+_x\chi_{Rx}) \right]
\right\} \ .
\ee
 Here $K$ is the fermion hopping parameter, $r$ the Wilson-parameter,
 which will be fixed to $r=1$ in the numerical simulations, and the
 indices $L,R$ denote, as usual, the chiral components of fermion
 fields.
 In this normalization the fermion-mirror-fermion mixing mass is
 $\mu_{\psi\chi}=1-8rK$.
 In the limit $\lambda \to \infty$ the length of the scalar field is
 frozen to unity, therefore in $S_\varphi$ only the term proportional
 to $\kappa$ is relevant.

 The consequence of the Golterman-Petcher identities is that
 at $G_\chi=0$ all higher vertex functions containing the $\chi$-field
 vanish identically, and the $\chi$-$\chi$ and $\chi$-$\psi$
 components of the inverse fermion propagator $\tilde{\Gamma}_\Psi(p)$
 are equal to the corresponding components of the free inverse
 propagator \cite{GOLREV,LINWIT}.
 In the broken phase the small momentum ($p \to 0$) behaviour of
 $\tilde{\Gamma}_\Psi$ defines the renormalized $\psi$-mass
 $\mu_{R\psi}$ and wave function renormalization factor $Z_\psi$,
 therefore
$$
\tilde{\Gamma}_\Psi(p) \equiv M + i \gamma \cdot p\, N + O(p^2)
$$
\be \label{eq03}
= \left(
\begin{array}{cc}
(\mu_{R\psi} + i\gamma \cdot \bar{p} + O(p^2))Z_\psi^{-1}  &
\mu_0 + \frac{r}{2}\hat{p}^2  \\[1em]
\mu_0 + \frac{r}{2}\hat{p}^2  &  i\gamma \cdot \bar{p}
\end{array}                            \right) \ .
\ee
 Here $\mu_0 \equiv \mu_{\psi\chi}/(2K) = (1-8rK)/(2K)$ and, as usual,
 $\bar{p}_\mu \equiv \sin p_\mu$ and
 $\hat{p}_\mu \equiv 2\sin \half p_\mu$.
 The propagator is the inverse of $\tilde{\Gamma}_\Psi$.
 With the notation
\be \label{eq04}
\mu_p \equiv \mu_0 + \frac{r}{2} \hat{p}^2
\ee
 we have
$$
\tilde{\Delta}_\Psi(p) = \tilde{\Gamma}_\Psi(p)^{-1}
\equiv A - i \gamma \cdot p\, B + O(p^2)
=  \left[ (\bar{p}^2 + Z_\psi\mu_p^2)^2 + \mu_{R\psi}^2 \bar{p}^2
   \right]^{-1}
$$
\be \label{eq05}
\cdot \left(
\begin{array}{cc}
Z_\psi [ \mu_{R\psi} \bar{p}^2 - i\gamma \cdot \bar{p}
(\bar{p}^2 + Z_\psi\mu_p^2) ]  &
Z_\psi\mu_p (\bar{p}^2 + Z_\psi\mu_p^2 + i\gamma \cdot \bar{p}\,
\mu_{R\psi})  \\[1em]
Z_\psi\mu_p (\bar{p}^2 + Z_\psi\mu_p^2 + i\gamma \cdot \bar{p}\,
\mu_{R\psi})  &
-Z_\psi\mu_p^2\mu_{R\psi} - i\gamma \cdot \bar{p}
(\bar{p}^2 + \mu_{R\psi}^2 + Z_\psi\mu_p^2)
\end{array}                            \right) + O(p^2) \ .
\ee

 This shows that in the broken phase near $p=0$ the elements of
 $\tilde{\Delta}_\Psi(p)$ are rapidly changing, unless $\mu_p$ is very
 small.
 Consider, for instance, $\tilde{\Delta}_{\psi\psi}$:
\be \label{eq06}
\tilde{\Delta}_{\psi\psi} = Z_\psi \frac{ \mu_{R\psi}
\left( 1 + Z_\psi \frac{\mu_p^2}{\bar{p}^2} \right)^{-1}
- i\gamma \cdot \bar{p} }
{ \bar{p}^2 + Z_\psi\mu_p^2 + \mu_{R\psi}^2
\left( 1 + Z_\psi \frac{\mu_p^2}{\bar{p}^2} \right)^{-1} } + O(p^2) \ .
\ee
 One sees that in the phase with broken symmetry ($\mu_{R\psi} \ne 0$)
 for $\mu_0$ and $p$ both being small there is a qualitative change
 of the behaviour as a function of $p$ around $p^2 \simeq \mu_0$.
 The limits $\mu_0 \to 0$ and $p \to 0$ cannot be interchanged.
 The correct order is to take first $\mu_0 \to 0$ and then $p \to 0$.
 Smooth behaviour near $\mu_0=0$ is reached only if
\be \label{eq07}
\mu_0 = O(p^2) \ .
\ee
 In this case $\mu_p^2/\bar{p}^2 = O(p^2)$, and the components of the
 propagator are
$$
\tilde{\Delta}_{\psi\psi}(p) = Z_\psi \frac{ \mu_{R\psi} -
i\gamma \cdot \bar{p} } { \mu_{R\psi}^2 + \bar{p}^2 } + O(p^2) \ ,
$$
\be \label{eq08}
\tilde{\Delta}_{\psi\chi}(p) = \tilde{\Delta}_{\chi\psi}(p)
= i\gamma\cdot\bar{p} \frac{\mu_p Z_\psi \mu_{R\psi}}
{\bar{p}^2 (\mu_{R\psi}^2 + \bar{p}^2)} = O(p) \ ,
\hspace{3em}
\tilde{\Delta}_{\chi\chi}(p) = \frac{-i\gamma \cdot \bar{p}}
{\bar{p}^2} + O(p^2) \ .
\ee
 This limit is best approximated if for a given momentum $p$ we choose
\be \label{eq09}
\mu_p = \mu_0 + \frac{r}{2} \hat{p}^2 = 0 \ .
\ee
 In fact, since our fermionic renormalized quantities are defined at
 the smallest timelike momentum $p_{min}$ (and zero spacelike momenta),
 we took
\be \label{eq10}
\mu_{p_{min}}=0 \ .
\ee
 Note that this also implies
 $\tilde{\Delta}_{\psi\chi}(p_{min}) = 0$, which ensures that no
 mixing between $\psi$ and $\chi$ occurs.
 On a lattice with time extension $T$ the smallest fermion momentum
 in the timelike direction with our antiperiodic boundary conditions
 is $p_{min}=\pi/T$, therefore this condition gives a $T$-dependent
 hopping parameter $K > K_{cr} \equiv 1/(8r)$.
 For $T \to \infty$ one has, of course, $K \to K_{cr}$.

 In summary: since in the broken phase the decoupling situation is
 reached only for very small $\mu_0$, in terms of the hopping parameter
 one has to be so close to $K=K_{cr} \equiv 1/(8r)$ that trying to
 perform simulations at $K < K_{cr}$ and to extrapolate to $K=K_{cr}$
 does not pay in the parameter region we studied (see in next section).
 Therefore in our simulations we always took
 $\mu_0 \simeq \mu_{p_{min}}=0$ ($K \simeq K_{cr}=0.125$).

 In numerical simulations one can determine the fermion propagator
 $\tilde{\Delta}_\Psi$.
 For the renormalized quantities one needs the inverse propagator
 $\tilde{\Gamma}_\Psi$.
 This could, in principle, be obtained for a given momentum by the
 numerical inversion of the $8 \otimes 8$ matrix in
 spinor-$\psi$-$\chi$ space, but this would introduce large
 fluctuations in the results.
 In \cite{FKLMMM,LMMW} we used an analytic inversion up to $O(p^2)$.
 Here it is better to take
$$
M = (A + \bar{p}^2 B A^{-1} B)^{-1} \ ,
$$
\be \label{eq11}
N = A^{-1} B M = A^{-1} B (A + \bar{p}^2 B A^{-1} B )^{-1} \ ,
\ee
 which follows from the spinor structure
 $\tilde{\Delta}_\Psi = A - i\gamma \cdot \bar{p}B$, and has at most
 $O(p^4)$ corrections.
 In addition, for the exactly known elements of $\tilde{\Gamma}_\Psi$
 we took the values in (\ref{eq03}), and (\ref{eq11}) was used only
 for the $\psi$-$\psi$ component of $\tilde{\Gamma}_\Psi$.

 The definition of the renormalized physical quantities can be
 taken over in most cases with trivial modifications from the
 $\rm U(1)_L \otimes U(1)_R$ model \cite{FKLMMM,LMMW}.
 In the broken phase, where most of our runs were performed, we also
 use the ``constraint correlations'' obtained after an SU(2)
 rotation of the average $\varphi$-field into a fixed (``$\sigma$'')
 direction.
 (The three Goldstone boson components of the $\varphi$-field
 perpendicular to $\sigma$ are denoted by $\pi_a\; (a=1,2,3)$.)
 In order to avoid infrared singularities, external $\pi$-legs are
 usually set to the smallest nonzero momentum on our $L^3 \cdot T$
 lattices.

 The renormalized Yukawa-couplings can be defined in different ways.
 One definition is given by the ratio of the mass to the renormalized
 vacuum expectation value, as for instance
 $G_{R\psi}=\mu_{R\psi}/v_R$.
 It is interesting to compare this to
 the renormalized Yukawa couplings obtained through
 Goldstone-fermion-antifermion vertex functions.
 These renormalized Yukawa couplings, denoted as $G^{(3)}_{aR\psi}$ and
 $G^{(3)}_{aR\chi}$, where $a=1,2,3$, are defined by
\be \label{eq12}
\left(
\begin{array}{cc}
 i\, \gamma_5\tau_a\, G^{(3)}_{aR\psi} & 0\\
  0      &-i\,\gamma_5\tau_a\, G^{(3)}_{aR\chi}
\end{array}
\right) \delta_{k,-p+q}
= {{\hat k}^2_4\over\sqrt{Z_T}}\,\tilde{\Gamma}_R(p_4)\, Z_\Psi^{-1/2}
\,G^{(c)}_{a}\, (Z_\Psi^{-1/2})^T\, \tilde{\Gamma}_R(q_4)\, \ ,
\ee
 where no summation over $a$ is applied, and $k_4$, $p_4$, $q_4$ are
 the $4$th components of the momenta of Goldstone boson, fermion and
 anti-fermion, respectively.
 We have set the spatial components of all momenta to zero.
 The appearance of the Kronecker-delta above is due to
 energy-momentum conservation.
 The renormalized 2-point fermion vertex function $\tilde{\Gamma}_R$
 at small $p=(0,0,0,p_4)$ is given as
$$
\tilde{\Gamma}_R(p_4) \simeq i\,\gamma_4\,{\bar p}_4\,+\, M_R\, ,
\hspace{3em}
   M_R = \left(
         \begin{array}{cc}
          G_{R\psi}v_R& \mu_R\\
          \mu_R        & G_{R\chi}v_R
         \end{array}
         \right)
\,\, ,
$$
and
$$
G^{(c)}_{a} = \frac{1}{L^3T}\,
\sum_{x,y,z}\,e^{-ik_4x_4}\,e^{-ip_4y_4}\,
e^{iq_4z_4}\Bigl<\pi_a(x)\Psi(y)\bar\Psi(z)\Bigr >_c
$$
 is the connected part of the $\pi_a$-$\Psi$-$\bar\Psi$ 3-point Green's
 function and $\pi_a(x)$ are the Goldstone fields
 ($\Psi(y) \equiv \Psi_y$ is the fermion field).
 Using the fact that the renormalized couplings are the same for all
 three Goldstone bosons, and
$$
{\rm Tr}_{Dirac}\,(\gamma_5^2) = 4\,  , \hspace{2em}
{\rm Tr}_{SU(2)}\,(\tau_a\tau_b)=2\delta_{ab}\,\, ,
$$
we obtain
$$
\left(
\begin{array}{cc}
   G^{(3)}_{R\psi}  &  0  \\
   0  & -G^{(3)}_{R\chi}
\end{array}
\right) \delta_{k,-p+q}
={-i{\hat k}^2_4\over {24\sqrt{Z_T}}}
$$
\be \label{eq13}
\cdot \sum_{a=1}^3\,{\rm Tr}_{SU(2)}\,\Bigl\{\tau_a\,
 {\rm Tr}_{Dirac}\,\bigl [\gamma_5\,
\tilde{\Gamma}_R(p_4)\, Z_\Psi^{-1/2}\, G^{(c)}_{a}\,(Z_\Psi^{-1/2})^T\,
\tilde{\Gamma}_R(q_4)\bigr ]\Bigr\}
\,\, .
\ee
 Because of the existence of massless Goldstone bosons in the broken
 phase, renormalized quantities cannot be defined at zero momentum.
 For instance, the connected 3-point $\pi$-$\Psi$-$\bar{\Psi}$
 Green's function has an infrared singularity on the external
 $\pi$-leg.
 Therefore in our simulations on $L^3 \cdot T$ lattices we choose
$$
k_4 = {2\pi\over T}\,\, ,\,\,
p_4 = -{\pi\over T}\,\, ,\,\, q_4={\pi\over T}
\,\, .
$$
 After carrying out all the matrix multiplications in (\ref{eq13}), we
 get $G^{(3)}_{R\psi}$ and $G^{(3)}_{R\chi}$.
 The expressions are too voluminous to be displayed here.


\section{ Numerical simulation }                            \label{s3}

 We used the Hybrid Monte Carlo algorithm \cite{DKPR}.
 This requires the flavour duplication of the fermion spectrum.
 If the fermion matrix in the action (\ref{eq02}) is denoted by $Q$,
 then the replica flavours have $Q^\dagger$, therefore for them the
 r\^oles of $\psi$ and $\chi$ are interchanged: $\chi$ is the
 ``fermion'' and $\psi$ the ``mirror fermion''.
 Since in the model under consideration the fermions are equivalent to
 mirror fermions, in the decoupling limit $G_\chi=\mu_{\psi\chi}=0$
 the model describes two degenerate fermion doublets
 (corresponding to the $\psi$-fields) and two massless ``sterile''
 doublets (belonging to the $\chi$'s), which have no interactions with
 the physical sector.

 The commonly used algorithm for the inversion of the fermion matrix
 is the conjugate gradient algorithm (CGA).
 Motivated by the paper of Gupta et al. \cite{GUPTA} we were also
 testing matrix inversion by minimal residual algorithm
 (MRA) with odd-even ($o$-$e$) decomposition in our SU(2) symmetric
 Higgs-Yukawa model with mirror pairs of fermion fields.
 One difference between our case and that of \cite{GUPTA} is that we
 have a nontrivial $Q_{ee}$ and $Q_{oo}$ instead of their matrix
 $M{\bf 1}$.
 But the requirements for a successful implementation of the MRA are
 satisfied in our case too.
 Namely we can invert $Q_{ee}$ and $Q_{oo}$ explicitly with an amount
 of time that is negligible in comparison with the algorithmic
 inversion of the full matrix $Q$.

 To compare the MRA with the CGA we solve the equation $Q^+ Qp=v$ for
 some scalar field configuration $\phi$ at different values of
 Yukawa-couplings, hopping parameter, lattice size and different
 convergence parameter.
 This last quantity is defined by:
\be \label{eq14}
\delta = \frac{|Q^+Qp-v|^2}{|v|^2} \ .
\ee
 Using the CGA the solution $p$ is accepted as soon as $\delta$ is
 smaller than some prescribed $\delta_0$.
 The solution by MRA is done in two steps, first solving $Q^+\bar{p}=v$
 and then $Qp=\bar{p}$, both with a bound for $\delta$ that is a
 factor of 100 smaller than $\delta_0$.

 We were also testing the ``polynomial preconditioning'' for the MRA,
 described in \cite{GUPTA}.
 Preconditioning of order $n$, denoted by MRA($n$), is characterized by
 the fact that the matrix, which is to be inverted, contains the
 fermion hopping parameter $K$ to the power $2n$.
 The tests were performed at $\lambda = 1.0$ and $\lambda = 10^{-6}$.
 No important differences were observed for different $\lambda$.
 As a few test runs showed, at $\lambda=\infty$ the algorithm behaves
 similarly to $\lambda = 1.0$.

\begin{table}[tb]
\caption{  \label{tb1}
 Comparison of matrix inversion algorithms for various $\delta_0$
and $G_{\chi}$.\newline
 $8^3 \cdot 16$ lattice, $G_{\psi}=0.1$, $\kappa=0.15$, $K=0.1$,
 $\lambda=1.0$ . }
\begin{center}
\begin{tabular}{|l||c|r|r|r|r|}
\hline
$\delta_0$ & $G_{\chi}$ & CGA & MRA(1) & MRA(2) & MRA(3) \\
\hline\hline
$10^{-8}$  &  0.0      &  6.5 &  3.6   &  4.5   &  5.5   \\
$10^{-15}$ &  0.0      & 12.6 &  6.1   &  7.0   &  8.5   \\
$10^{-20}$ &  0.0      & 16.8 &  7.9   &  8.9   & 10.7   \\
$10^{-25}$ &  0.0      & 21.1 &  9.7   & 10.7   & 12.6   \\
$10^{-30}$ &  0.0      & 25.2 & 11.8   & 12.5   & 14.9   \\
$10^{-8}$  & -0.1      &  5.8 &  3.5   &  4.1   &  5.1   \\
$10^{-15}$ & -0.1      & 10.5 &  5.5   &  6.5   &  7.7   \\
$10^{-30}$ & -0.1      & 20.5 & 10.4   & 11.1   & 13.0   \\
$10^{-8}$  &  0.2      & 11.4 &  5.0   &  6.1   &  7.8   \\
$10^{-15}$ &  0.2      & 24.4 &  8.8   & 10.2   & 12.2   \\
$10^{-30}$ &  0.2      & 52.3 & 17.6   & 19.0   & 22.0   \\
\hline
\end{tabular}
\end{center}
\end{table}
\begin{table}[tb]
\caption{  \label{tb2}
 Comparison of matrix inversion algorithms for various $K$.\newline
 $4^3 \cdot 8$ lattice,
 $\delta_0=10^{-8}$, $G_{\psi}=0.3$, $G_{\chi}=0.0$, $\kappa=0.09$,
 $\lambda=10^{-6}$ .  }
\begin{center}
\begin{tabular}{|l||c|c|c|c|}
\hline
K     & CGA   & MRA(1)  & MRA(2)  & MRA(3)   \\
\hline\hline
0.10  & 0.59  & 0.24    & 0.30    & 0.37     \\
0.11  & 0.87  & 0.35    & 0.41    & 0.59     \\
0.12  & 1.51  & 0.91    & 1.14    & 1.41     \\
0.123 & 1.70  & 2.41    & 2.87    & 2.80     \\
0.124 & 1.74  & 4.37    & 4.45    & 3.90     \\
\hline
\end{tabular}
\end{center}
\end{table}
\begin{table}[tb]
\caption{  \label{tb3}
 Comparison of different preconditionings at $K=0.125$.\newline
 $4^3\cdot 8$ lattice,
 $\delta_0=10^{-8}$, $G_{\psi}=0.3$, $G_{\chi}=0.0$, $\kappa=0.09$,
 $\lambda=10^{-6}$ .  }
\begin{center}
\begin{tabular}{|c|c|c|c|c|c|c|c|}
\hline
CGA & \multicolumn{7}{c|}{MRA(n)} \\
\cline{2-8}
    & n=1 & 2 & 3 & 4 & 5 & 6 & 7 \\
\hline\hline
1.87  & 11.30   & 8.58    & 6.11    & 3.95    & 2.29    & 12.90   &
15.00 \\
\hline
\end{tabular}
\end{center}
\end{table}
\begin{table}[tb]
\caption{  \label{tb4}
 Comparison of matrix inversion algorithms on different lattice
 sizes\newline
 $\delta_0=10^{-15}$, $G_{\psi}= 0.1$, $G_{\chi}=0.0$, $\kappa=0.15$,
  $K=0.1$, $\lambda=1.0$ .  }
\begin{center}
\begin{tabular}{|c||r|c|c|}
\hline
$L^3 \cdot T$ & CGA  & MRA(1) & MRA(2)  \\
\hline\hline
$4^3 \cdot 8 $ & 0.67  & 0.41   & 0.46    \\
$6^3 \cdot 12$ & 3.84  & 1.97   & 2.24    \\
$8^3 \cdot 16$ & 12.60 & 6.10   & 7.00    \\
\hline
\end{tabular}
\end{center}
\end{table}

 Our results are summarized in tables \ref{tb1} to \ref{tb4},
 where the CPU time necessary for solving $Q^+Qp=v$ is given in seconds.
 In these tests we could not find any gain using overrelaxation.

 Table \ref{tb1} shows for $K=0.1$ that the smaller $\delta_0$ becomes,
 the better is the MRA compared to CGA.
 Comparing different preconditionings MRA(1) is the best.
 This picture changes, if one looks at higher $K$ as can be seen from
 tables \ref{tb2} and \ref{tb3}.
 Near the critical value of $K$ simple preconditioning is no longer the
 best choice and even with an optimal preconditioning of MRA the CGA
 performs better.

 Table \ref{tb4} shows a comparison between MRA and CGA at $K=0.1$ for
 different lattice sizes.
 One observes that the gain of using MRA increases with the lattice
 size.

 Our conclusion is that in the case of small $K$ ($K \leq 0.1$) the MRA
 has to be preferred.
 For some choices of parameters, e.g.\ in the symmetric phase, the gain
 by using MRA can be so large that it would be advantageous to do
 calculations at different small values of $K$ (e.g. $0.1,0.11,0.12$)
 and then extrapolate to the value under investigation (e.g. $0.125$).
 However for $K$ deep in the broken phase the CGA seems to be still the
 best choice.
 For the investigation of the decoupling limit ($K \simeq K_{cr}=0.125$)
 in the broken phase it was necessary to be very close to $K_{cr}$, and
 we always used CGA.

 The technical advantage of using the decoupling method is that the
 number of tuned parameters is less, because the fermion hopping
 parameter is fixed at $K=K_{cr}$.
 At $\lambda=\infty$ and at a fixed value of the bare Yukawa-coupling
 $G_\psi$ one has to tune only the scalar hopping parameter $\kappa$.
 The difficulty is that the presence of massless $\chi$-fermions
 slows down the convergence of the fermion matrix inversion.
 We also tried to improve on this by using the free fermion propagator
 for preconditioning in momentum space.
 In this way the number of iterations in CGA can be reduced by a factor
 not larger than two.
 On the other hand, the computer time required for performing the
 necessary Fourier transformations is so large that the gain is
 completely counteracted, except for very small Yukawa-couplings.
 This is presumably due to the ``roughness'' of typical scalar field
 configurations.

 Another attempt to improve the matrix inversion was to supply the CGA
 with an ``educated guess'' for the start vector in the iterations.
 This was done by means of a hopping parameter expansion of $Q^{-1}$ up
 to some order $n$:
$$
Q^{-1} \approx
Q^{-1}_{n} = D^{-1} \left( 1 - \sum_{k=1}^n (M D^{-1})^k \right) \ ,
$$
 where $D$ and $M$ are the diagonal and off-diagonal parts of $Q$
 with respect to site indices, and $n$ was optimized for given
 parameters $K$ and $G_{\psi}$.
 For $K=0.1$ and $G_{\psi} < 2.4$ the speed of the algorithm can be
 increased up to a factor of three in this way.
 On the other hand, for $K$ above its critical value this improvement is
 not applicable.


\section{ Results }                                         \label{s4}

 The first step in the numerical simulations was to check the
 phase structure at $\lambda=\infty$ and $K = K_{cr}$.
 On the basis of experience with several other lattice Yukawa models
 \cite{GOLREV,SHIREV}, and our own previous work \cite{LIMOWI}, this is
 expected to possess several phase transitions between the
 {\it ``ferromagnetic''} (FM), {\it ``antiferromagnetic''} (AFM),
 {\it ``paramagnetic''} (PM) and {\it ``ferrimagnetic''} (FI) phases.
 The resulting picture in the ($G_\psi,\kappa$)-plane is shown in
 fig.~1.
 Due to CPU-time limitations we did not try to disentangle the details
 of the structure near the meeting point of the four phases, neither did
 we follow the shape of the FI-phase for very strong bare
 Yukawa-coupling beyond $G_\psi=1.5$.

 The physical phase is FM with spontaneously broken chiral symmetry.
 Therefore we fixed $G_\psi=0.3,\; 0.6,\; 1.0$ and performed a series
 of runs in the $\kappa$-ranges shown in fig.~1 by the dashed lines.
 Most of the time $4^3 \cdot 8$ and $6^3 \cdot 12$ lattices were
 taken.
 In a few particularly important points, for instance at
 ($\kappa=0.27,\, G_\psi=0.3$) and ($\kappa=0.15,\, G_\psi=0.6$),
 in addition to $4^3 \cdot 8$ and $6^3 \cdot 12$ also an $8^3 \cdot 16$
 run was performed.
 The typical run consisted of about 1000-2000 equilibrating and
 4000-10000 measured HMC trajectories.
 The length of trajectories was randomly changed by the number of
 classical dynamics steps between 3 and 10.
 The step length was chosen such that the average acceptance rate per
 trajectory stayed near 0.75.

 On our lattices the expectation value of the average of the scalar
 field in the $\sigma$-direction
 $v \equiv \langle \sigma_x \rangle = \langle \phi_{Lx} \rangle$
 (in short ``magnetization'') has a smooth behaviour across the
 physically important PM-FM phase transition (part of our data is
 shown in fig.~2).
 Furthermore, the magnetization always decreased with increasing
 lattice size, in the same way as in case of the pure O(4)-symmetric
 $\phi^4$ model.
 This agrees with the expected second order phase transition, which
 is well suited for the definition of a continuum limit.

 In all data points the mass of the mirror fermion, $\mu_{R\chi}$
 and the renormalized mixing mass, $\mu_R$ were consistent with zero
 within errors, in agreement with the consequences of the
 Golterman-Petcher relations.

 The behaviour of the fermion mass $\mu_{R\psi}$ and Higgs-boson mass
 in lattice units is shown in figs.~3 and 4, respectively.
 The fermion mass ($\mu_{R\psi}$) is decreasing monotonically to zero,
 as one approaches the phase transition from the FM-side (decreasing
 $\kappa$).
 This agrees with the expectation, since in the PM phase, due to
 $K \simeq K_{cr}$, the fermion mass is nearly zero (at $K=K_{cr}$ on an
 infinite lattice it would be exactly zero).
 At the same time the fermion mass is increasing with $G_\psi$, in
 such a way that at $G_\psi=1.0$ within our limited computer time
 we were unable to find points with really small masses.
 This could presumably be cured by investigating more points on
 larger lattices.
 In fact, the fermion mass shows in this point strong finite size
 effects, implying smaller masses on larger lattices.

 The masses of the doubler fermions for both $\psi$ and $\chi$ at
 nonzero corners of the Brillouin-zone were also determined and turned
 out to be always above 1.5, with a slight decreasing tendency for
 increasing $G_\psi$.

 The values of the Higgs-boson mass ($m_{R\sigma} \equiv m_L$)
 were determined by fitting the constraint correlation in the
 $\sigma$-channel by a form $\cosh() + const.$ in the range
 $1 \le t \le T/2$.
 The dependence of $m_{R\sigma}$ on $\kappa$ in fig.~4 shows a gradual
 decrease and then a sharp increase for decreasing $\kappa$.
 On larger lattices the values are smaller, but there is a substantial
 increase with increasing $G_\psi$ if the lattice size is kept fixed.
 This is depicted in fig.~5, where the averages of a few points with
 lowest Higgs-boson mass are shown.
 This figure also displays the strong finite size effects present
 on these lattices: in the limit of infinitely large volumes the
 minimum of $m_{R\sigma}$ is expected to be zero at the second order
 phase transition between the FM-PM phases.
 A plausible interpretation of fig.~5 is that the finite size effects
 become stronger for larger $G_\psi$, because the renormalized
 couplings become stronger.
 In any case, the large values at the minima represent a difficulty
 for the numerical simulations in the critical region, because large
 lattices are needed.
 Our experience shows (see also table~\ref{tb5}), that reasonably small
 masses $m_{R\sigma} \simeq 0.5-0.7$ can be achieved at $G_\psi=0.3$ on
 $6^3 \cdot 12$, at $G_\psi=0.6$ on $8^3 \cdot 16$ lattices.
 Presumably at $G_\psi=1.0$ lattices with spatial extension of
 at least $16^3$ are necessary.
 In general, for a physical interpretation of the results in the broken
 phase on a lattice of given size one should stay with $\kappa$ above
 the value where $m_{R\sigma}$ takes its minimum.
 This expectation is strengthened by the comparison of $6^3 \cdot 12$
 and $8^3 \cdot 16$ results at ($\kappa=0.27,\, G_\psi=0.3$), which
 are both in the broken phase, and within statistical errors show
 no finite size effects of the renormalized couplings (see below).

 The behaviour of the $\sigma$- (Higgs-boson) and $\pi$-
 (Goldstone-boson) inverse propagators as a function of momentum is
 shown at $G_\psi=0.6,\kappa=0.15$ in fig.~6.
 The method of measurement is the same as in Ref.~\cite{BDFJT}.
 The Goldstone- ($\pi$-) and $\sigma$-propagators in momentum space
 are defined by
\begin{eqnarray}
\tilde{G}_{\pi}(p) &=& \left\langle \frac{1}{3L^3} \sum_{x,y}
\sum_{a=1}^{3} \pi_{ax} \pi_{ay}
\exp\{ i p \cdot (x-y) \} \right\rangle \; ,
\label{eq15} \\
\tilde{G}_{\sigma}(p) &=& \left\langle \frac{1}{L^3} \sum_{x,y}
\sigma_x \sigma_y
\exp\{ i p \cdot (x-y) \} \right\rangle \; .
\label{eq16}
\end{eqnarray}
 In order to limit computer time and storage we actually measure
 $\tilde{G}(p)$ for only one of multiple 4-momenta $p$ giving degenerate
 $\hat{p}^2$ according to the assumption that, at least for small
 momenta, $\tilde{G}(p)$ is just a function of $\hat{p}^2$.
 The values are blocked during the MC runs with a block length
 of typically 100 configurations.
 The error bars are estimated with the jackknife method.
 As one can see, in this point the inverse $\pi$-propagator extrapolates
 reasonably well to zero for zero momentum.
 The extrapolation of the inverse $\sigma$-propagator to zero gives
 a Higgs-boson mass $m_{R\sigma}=0.98 \pm 0.05$, in good agreement with
 the value obtained from a fit of the time-dependence of timeslices.
 The curvature of the inverse propagators at this $(\kappa,G_\psi)$
 value is not strong.
 This allows a reasonably accurate determination of the renormalized
 quantities by the formulae in \cite{LMMW}, assuming a linear
 dependence between zero and the lowest nonzero momentum.
 For larger values of the momentum the propagators are quite smooth,
 therefore the effect of heavy $\psi$-fermion doublers is not strong.
 Closer to the phase transition the curvature of the inverse propagators
 near zero momentum becomes stronger, and the $\pi$-propagator starts
 to show an increasingly nonzero mass.
 We interpret the latter as a finite size effect.

 Taking the Goldstone-boson field renormalization factor
 $Z_\pi \equiv Z_T$ from the $\pi$-propagator, one can determine the
 renormalized scalar vacuum expectation value
 $v_R=\langle \sigma_x \rangle /\sqrt{Z_\pi}$, which in turn gives the
 renormalized quartic- ($g_R$) and Yukawa-couplings ($G_{R\psi}$) by
\be \label{eq17}
g_R \equiv \frac{3m_{R\sigma}^2}{v_R^2} \ ,  \hspace{4em}
G_{R\psi} \equiv \frac{\mu_{R\psi}}{v_R} \ .
\ee
 These are shown in figs.~7 and 8.
 Although the errors are quite large, one can observe a strong
 increase of $g_R$ for decreasing $\kappa$.
 In the $\kappa$ region above the minimum of $m_{R\sigma}$ there is
 much less variation.
 The values on these ``plateaus'' show a moderate increase for
 increasing $G_\psi$.
 The renormalized Yukawa-coupling $G_{R\psi}$ is rather flat as a
 function of $\kappa$, but increases definitely with $G_\psi$.

 Considering only the points above the minimum of $m_{R\sigma}$ on the
 given lattice size as being in the broken phase, one can make a
 tentative first estimate of the upper bound on the renormalized
 quartic coupling (or Higgs-boson mass), as a function of the
 renormalized Yukawa-coupling (or fermion mass).
 Such an estimate is shown by fig.~9 together with the perturbative
 estimates based on the 1-loop $\beta$-functions.
 The agreement with the perturbative results at $G_\psi=0.3$ and
 0.6 is good, although the renormalized couplings are quite strong,
 i.~e. close to the tree unitarity bound.
 The $8^3 \cdot 16$ points have, unfortunately, larger statistical
 errors.
 At strong coupling the good agreement could partly be due to a fixed
 point in the ratio $g_R/G_{R\psi}^2$, which implies that this ratio is
 insensitive to the cut-off.

 A few measured physical quantities in selected typical points are
 collected in table \ref{tb5}.
 Comparing the results at $G_\psi=0.6,\kappa=0.15$ on $4^3 \cdot 8$,
 $6^3 \cdot 12$ and $8^3 \cdot 16$ lattices with label b, c and d,
 respectively, one can see the evolution of the finite size effects.
 Between c and d there is much less change than between b and c, but
 point d on the $8^3 \cdot 16$ lattice still does somewhat differ
 from the infinite volume limit.
 This may explain why the corresponding point in fig.~9 is higher
 than the $6^3 \cdot 12$ points at larger $\kappa$.
 The situation is better if one compares points C and D, where the
 deviation of the renormalized couplings is within statistical errors.
\begin{table}[tb]
\caption{  \label{tb5}
 The main renormalized quantities and the bare magnetization
 $\langle \sigma \rangle \equiv \langle|\varphi|\rangle$ for several
 bare couplings $G_\psi$, and $\kappa$-values near the minimum scalar
 mass attainable for the given lattice size. Points labelled by capital
 letters are at $G_\psi=0.3$, whereas lower case and greek letters
 denote data obtained for $G_\psi=0.6$ and $G_\psi=1.0$, respectively. }
\begin{center}
\begin{tabular}
{||c|r@{$\cdot$}l|r@{.}l|r@{.}l|r@{.}l|r@{.}l|r@{.}l|r@{(}l|r@{.}l||}
\hline
\hline
&$L^3$&$\,T$ & \multicolumn{2}{c|}{$\kappa$}
&\multicolumn{2}{c|}{$\langle\sigma\rangle$} & \multicolumn{2}{c|}{$v_R$}
&\multicolumn{2}{c|}{$m_{R\sigma}$} &\multicolumn{2}{c|}{$\mu_{R\psi}$}
&\multicolumn{2}{c|}{$g_R$} & \multicolumn{2}{c||}{$G_{R\psi}$} \\
\hline
 A & $4^3$&$\, 8$ &  0&24 & 0&2807(15) & 0&307(4) & 1&23(1)
   & 0&34(2) &  48&3)   &  1&09(7) \\
 B & $6^3$&$\,12$ &  0&24 & 0&146(6)   & 0&18(1)  & 0&73(5)
   & 0&21(4) &  53&16)  &  1&2(4)  \\
 C & $6^3$&$\,12$ &  0&27 & 0&303(3)   & 0&309(13)& 0&80(5)
   & 0&39(2) &  20&4)   &  1&25(5) \\
 D & $8^3$&$\,16$ &  0&27 & 0&270(2)   & 0&25(1)  & 0&77(3)
   & 0&342(2)&  31&4)   &  1&35(6) \\
 E & $6^3$&$\,12$ &  0&30 & 0&4391(14) & 0&400(13)& 1&17(7)
   & 0&55(2) &  26&7)   &  1&36(6) \\
\hline
 a & $8^3$&$\,16$ &  0&12 & 0&118(5)   & 0&136(15)& 0&63(8)
   & 0&61(2) &  80&50)  &  4&5(3)  \\
 b & $4^3$&$\, 8$ &  0&15 & 0&3358(16) & 0&361(8) & 1&60(6)
   & 1&9(3)  &  59&7)   &  4&9(5)  \\
 c & $6^3$&$\,12$ &  0&15 & 0&248(2)   & 0&25(1)  & 1&14(5)
   & 0&67(6) &  63&10)  &  2&7(3)  \\
 d & $8^3$&$\,16$ &  0&15 & 0&218(3)   & 0&217(17)& 0&86(6)
   & 0&54(4) &  52&9)   &  2&5(3)  \\
 e & $6^3$&$\,12$ &  0&18 & 0&3524(18) & 0&36(2)  & 1&23(8)
   & 0&86(8) &  36&6)   &  2&4(3)  \\
 f & $6^3$&$\,12$ &  0&21 & 0&4390(17) & 0&41(2)  & 1&34(8)
   & 1&11(3) &  32&5)   &  2&71(13)\\
\hline
$\alpha$
   & $6^3$&$\,12$ & $-0$&12 & 0&189(2) & 0&243(14)& 1&79(15)
   & 0&95(9) & 180&40)  &  3&9(4)  \\
\hline
\hline
\end{tabular}
\end{center}
\end{table}

 A qualitative relation of the masses at the strongest Yukawa-coupling
 ($G_\psi = 1.0$) is that the mass of the Higgs-boson $m_{R\sigma}$
 is roughly twice as large as the fermion mass $\mu_{R\psi}$.
 Because of finite size effects and limited statistics it cannot be
 decided at present whether the $\sigma$-particle is a two-fermion bound
 state or a resonance near threshold.
 In the latter case, due to the fast decay into a fermion pair, the
 physical Higgs-boson could become a very broad resonance.

 An interesting question is the behaviour of the renormalized
 Yukawa-coupling $G^{(3)}_{R\psi}$ defined by the 3-point vertex
 function in (\ref{eq13}).
 $G^{(3)}_{R\psi}$ is smaller than $G_{R\psi}$ in all points.
 The measured values on $6^3 \cdot 12$ lattice are, for instance,
 $G^{(3)}_{R\psi}=1.0 \pm 0.6$ at point c and
 $G^{(3)}_{R\psi}=1.8 \pm 0.5$ at point $\alpha$.
 Therefore the ratio
\be \label{eq18}
S_3 \equiv \frac{G^{(3)}_{R\psi}}{G_{R\psi}}
\ee
 is smaller than 1.
 On the $8^3 \cdot 16$ lattice, within our statistics
 $G^{(3)}_{R\psi}$ turned out to be difficult to measure.
 An exception is point D where we obtained
 $G^{(3)}_{R\psi}=0.74 \pm 0.05$.
 This is also smaller by about a factor of 2 than the corresponding
 value of $G_{R\psi}$.
 The deviation could partly be due to the nonzero momentum value
 where $G^{(3)}_{R\psi}$ was extracted.

 The measured values of $G_{R\chi}$ were always consistent with zero
 within small errors, in agreement with the consequences of the
 Golterman-Petcher relations.


\section{ Conclusions }                                     \label{s5}

 The important trends seen in our numerical data on $4^3 \cdot 8$,
 $6^3 \cdot 12$ and $8^3 \cdot 16$ lattices are the following:
\begin{itemize}
\item
 The phase structure at $(\lambda=\infty,\, K=K_{cr})$ is qualitatively
 the same as in other lattice Yukawa-models with FM, PM, AFM and
 FI phases (fig.~1).
\item
 The FM-PM phase transition at $\lambda=\infty$ is smooth, probably of
 second order (fig.~2).
\item
 On most of our lattices there are strong finite size effects.
 In particular, large lattices are needed in order to bring the minimum
 of the Higgs-boson mass on a given lattice size down to interesting
 values below 1.
 As our simulations show, at moderate values of the bare
 Yukawa-coupling $8^3 \cdot 16$ might be enough, but for large values
 near $G_\psi \ge 1.0$ one will need at least presumably something like
 $16^3 \cdot 32$.
\item
 Considering only the $\kappa$-values above the minimum of the
 Higgs-boson mass on a given lattice size, where in one point we also
 have evidence that finite size effects are not very strong, we
 obtained a first tentative estimate of the upper bound on the
 renormalized quartic coupling as a function of the renormalized
 Yukawa-coupling (fig.~9).
 Up to renormalized Yukawa-couplings at the tree unitarity limit,
 which is reached near $G_\psi=0.6$, this agrees well with 1-loop
 perturbation theory, but further investigations are necessary in order
 to check finite size effects and extend the results towards larger
 $G_\psi$.
\item
 At the strongest Yukawa-coupling $G_\psi=1.0$ the mass of the physical
 Higgs-boson is roughly equal to twice the heavy fermion mass.
 This could imply that the Higgs-boson is a very broad resonance,
 which decays very fast into a heavy fermion pair.
\item
 For a given lattice size the renormalized Yukawa-coupling $G_{R\psi}$
 defined by the fermion mass increases more or less linearly with
 $G_\psi$ up to $G_\psi=1.0$, where it becomes almost twice the tree
 unitarity bound $\simeq 2.5$.
 $G^{(3)}_{R\psi}$ defined by the 3-point vertex function is smaller
 than $G_{R\psi}$ on the $6^3 \cdot 12$ and $8^3 \cdot 16$ lattices.
 As discussed above, the finite size effects are particularly strong
 for $G_\psi > 0.6$, therefore large lattices are needed for
 confirmation of the values of $G_{R\psi}$.
\end{itemize}

 The question of the possible influence of heavy fermions in the
 Standard Model is important and very interesting.
 By numerical simulations at $\lambda=\infty$ one can obtain
 information on the upper limit on the Higgs-boson mass.
 The extension to smaller values of the bare quartic coupling, in
 particular to $\lambda \simeq 0$ gives a lower bound related to
 vacuum stability.

 {\em Note added:} In writing this paper we received a recent preprint
 of Bock, Smit and Vink, where the same continuum ``target'' theory
 as ours has been numerically investigated in a staggered fermion
 formulation \cite{BOSMVI}.
%

\vspace{1cm}
\large\bf Acknowledgements \normalsize\rm \newline
\vspace{3pt}

\noindent
 We thank Jiri Jers\'ak for active help and useful discussions.
 The Monte Carlo calculations for this work have been performed
 on the CRAY Y-MP/832 of HLRZ J\"ulich and the S-600 of RWTH Aachen.

%
%
%
%
\newpage
\newpage
\begin{center}     \Large\bf Figure captions \normalsize\rm
\end{center}

\vspace{15pt}
\bf Fig.\,1.    \hspace{5pt} \rm
Phase structure of the $\rm SU(2)_L\otimes SU(2)_R$ symmetric Yukawa
model at $\lambda=\infty$ in the ($G_\psi,\,\kappa$)-plane. The
remaining bare parameters are fixed by the conditions $\mu_p=0,\,
G_\chi=0$. Open circles denote points in the PM~phase, crosses represent
points in the FM~phase. The points in the AFM and FI~phases are denoted
by full circles and open squares, respectively. The dashed lines
labelled~R,S,T each show the range of~$\kappa$ used for a systematic
scan of renormalized parameters at fixed~$G_\psi$. The crosses along
those lines denote the $\kappa$ values where the minimum scalar mass in
the broken phase is encountered. Solid lines connect the critical
values for $\kappa$ estimated from the behaviour of
$\langle\sigma\rangle^2$ on $4^3\cdot8$. Dashed lines around
the FI~phase show the expected continuation of the critical lines.

\vspace{15pt}
\bf Fig.\,2.    \hspace{5pt} \rm
The square of the magnetization as a function of $\kappa$. Here and in
the following figures, points at $G_\psi=0.3$ are represented by
triangles, points at $G_\psi=0.6$ by squares and points at $G_\psi=1.0$
by circles. Open symbols denote the $4^3\cdot8$ lattice, whereas
filled-in symbols stand for points obtained on $6^3\cdot12$.

\vspace{15pt}
\bf Fig.\,3.    \hspace{5pt} \rm
The fermion mass $\mu_{R\psi}$ plotted versus $\kappa$ for $G_\psi=0.3$
(triangles) and $G_\psi=0.6$ (squares) on lattices of size $4^3\cdot8$
(open symbols) and $6^3\cdot12$ (filled-in symbols). Errorbars are
omitted when the variation is of the size of the symbols. It is seen
that larger bare couplings $G_\psi$ in general yield larger fermion
masses.

\vspace{15pt}
\bf Fig.\,4.    \hspace{5pt} \rm
The scalar mass $m_{R\sigma}$ ploted versus $\kappa$. The explanation
of symbols corresponds to fig.~3. When approaching the phase transition
the scalar masses increase again after going through a minimum.

\vspace{15pt}
\bf Fig.\,5.    \hspace{5pt} \rm
The minimum of the scalar masses for different~$G_\psi$ on $4^3\cdot8$
(open symbols) $6^3\cdot12$ (filled-in symbols) and $8^3\cdot16$
(square plus vertex).
The $G_\psi$-values of the two points at $G_\psi=1.0$ are
slightly shifted in the plot to give a better separation.

\vspace{15pt}
\bf Fig.\,6a.   \hspace{5pt} \rm
The inverse propagator for the massive scalar field (open squares) and
the massless components (crosses) plotted versus the square of the
lattice momentum~$\widehat{p}$. The observed curvature is caused by
the interaction with fermions.

\vspace{15pt}
\bf Fig.\,6b.   \hspace{5pt} \rm
The inverse $\pi$-propagator from fig.~6a for the first few lattice
momenta.
The inverse of the renormalization constant~$Z_T$ is in principle
determined by the slope of the curve through the origin, and is
approximated by the slope of the straight line through the origin
and the point with smallest nonzero momentum, as shown by the dotted
line.

\vspace{15pt}
\bf Fig.\,6c.   \hspace{5pt} \rm
The inverse $\sigma$-propagator from fig.~6a for the first few lattice
momenta. Extrapolating the curve to zero momentum gives an estimate
for $m_{R\sigma}^2$.

\vspace{15pt}
\bf Fig.\,7.    \hspace{5pt} \rm
The renormalized Yukawa-coupling $G_{R\psi}$ versus $\kappa$ for
different bare $G_\psi$. The explanation of symbols is the same as in
fig.~2.

\vspace{15pt}
\bf Fig.\,8.    \hspace{5pt} \rm
The renormalized quartic coupling $g_R$ as a function of $\kappa$ for
three different bare values of $G_\psi$.

\vspace{15pt}
\bf Fig.\,9.    \hspace{5pt} \rm
The renormalized quartic coupling $g_R$ plotted versus $G_{R\psi}^2$.
Full data points are from runs on $6^3 \cdot 12$ and represent
the mean values obtained from the data points C,E and e,f in
table \ref{tb5}, respectively.
In addition, two runs on $8^3\cdot16$ are shown, namely point D (open
triangle plus vertex) and point d (open square plus vertex).
The solid and dotted curves show the results for the upper bound on
$g_R$ computed from the integration of the 1-loop $\beta$-functions for
a scale ratio $\Lambda/m_{R\sigma}=3,\,4$, respectively.
The first value for $\Lambda/m_{R\sigma}$ corresponds to
$m_{R\sigma} \simeq 1$ in lattice units, whereas the latter is
equivalent to $m_{R\sigma} \simeq 0.75$.
(The cut-off $\Lambda$ is defined to be $\pi$ in lattice units.)

\end{document}